\documentclass[12pt]{spieman}  
\usepackage{amsmath,amsfonts,amssymb}
\usepackage{graphicx}
\usepackage{setspace}
\usepackage{lineno}

\definecolor{updatecolor}{rgb}{0.5,0.0,0.8}
\newcommand{\updated}[1]{{#1}}

\title{The STROBE-X Low Energy Modular Array (LEMA) Instrument}

\author[a,*]{Keith C. Gendreau}
\author[b]{Dominic Maes}
\author[c]{Ronald A. Remillard}
\author[d]{Paul S. Ray}
\author[a]{Zaven Arzoumanian}
\author[a]{Craig Markwardt}
\author[a]{Takashi Okajima}
\affil[a]{NASA's Goddard Space Flight Center, Greenbelt, MD 20771 USA}
\affil[b]{BAE Systems Space \& Mission Systems Inc.}
\affil[c]{Kavli Institute for Astrophysics and Space Research, Massachusetts Institute of Technology, Cambridge MA 02139 USA}
\affil[d]{Space Science Division, U.S. Naval Research Laboratory, Washington, DC 20375 USA}
\begin{document} 
\maketitle


\begin{abstract}
 The Low Energy Modular Array (LEMA) is one of three instruments that compose the STROBE-X mission concept. The LEMA is a large effective-area, high throughput, non-imaging pointed instrument based on the X-ray Timing Instrument of the Neutron star Interior Composition Explorer (NICER) mission. The LEMA is designed for spectral-timing measurements of a variety of celestial X-ray sources, providing a transformative increase in sensitivity to photons in the 0.2--12 keV energy range compared to past missions, with an effective area (at 1.5 keV) of 16,000 cm$^2$ and an energy resolution of 85 eV at 1 keV.

\end{abstract}

\keywords{X-ray, probes, STROBE-X}

{\noindent \footnotesize\textbf{*}\linkable{keith.c.gendreau@nasa.gov} }

\section{Introduction}
\label{sect:intro}  

\begin{figure}[h]
    \centering
    \includegraphics{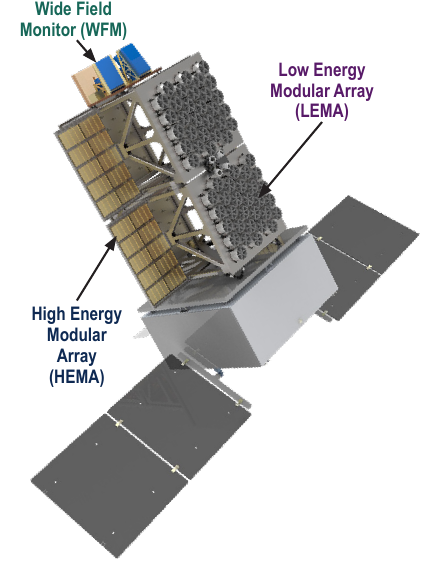}
    \caption{Rendering of the STROBE-X observatory, showing its three instruments.}
    \label{fig:overview}
\end{figure}

We present an overview of the Low Energy Modular Array (LEMA) Instrument on the \textit{Spectroscopic Time-Resolving Observatory for Broadband Energy X-rays\/} (STROBE-X) mission, which has been proposed in response to NASA's 2023 call for Probe-class X-ray missions. STROBE-X carries a suite of three instruments (see Fig.~\ref{fig:overview}) that work together to provide situational awareness of the X-ray sky and characterize the broadband spectral and timing behavior of sources from 0.2 to 50 keV\cite{JATISOverview}. 

\section{Instrument Overview}

The concept and design of the LEMA is based on the X-ray Timing Instrument (XTI) on the NICER Mission of Opportunity currently aboard the International Space Station \cite{2016SPIE.9905E..1HG}, offering an unprecedented level of scientific performance by providing over 8.5$\times$ the XTI collecting area with substantially reduced background in the 0.2--12 keV band. The hardware evolution from NICER to LEMA is achieved with four low-risk modifications: \updated{increased} size of the X-ray ``concentrator'' optics (XRCs), a composite optical bench, FPGA-based onboard event processing (replacing obsolete microcontrollers), and improved control of detector resets. As its name implies, the LEMA is highly modular in design, providing the instrument with inherent resilience (no single point of failure and graceful degradation), parallelism in manufacturing/testing/integration, and 
enhanced thermal-mechanical stability of optical alignment within and across modules.
The instrument consists of an aligned collection of 60 concentrator optics and commercially available silicon drift detectors (SDDs), both with deep heritage from the NICER instrument. The optic-detector pairs are split into two groups of 30, which are arranged on identical composite optical benches. Each XRC collects photons from a large geometric area over a 20 arcmin$^2$ region of the sky and concentrates them onto a small SDD. The SDD detects individual X-ray photons, recording their energies and times of arrival to high precision. Together, this assemblage provides a photon counting capability with large effective area, high time resolution, moderate energy resolution, and low background.  Table \ref{tab:inst} summarizes the main characteristics of the LEMA instrument.

\begin{table}[h]
\caption{LEMA Parameters\label{tab:inst}}
\begin{center}
\begin{tabular}{ll}
\hline
\textbf{Instrument Characteristic} & \textbf{Value (CBE)} \\
\hline\hline
Bandpass & 0.2--12 keV \\
Collecting Area @ 1.5 keV  & 16,000 cm$^2$ \\
Non-Imaging FoV  & 4 arcmin (FWHM) \\
Count Rate on 1   Crab &  111,000 c/s\\
Background Level & \updated{3 c/s} ($6 \times 10^{-13}$ erg/cm$^2$/s)\\
Sensitivity (5$\sigma$ in 1 ks Crab-like spectrum) & $1.5 \times 10^{-13}$ erg/cm$^2$/s \\
Timing Resolution & 40 ns \\
Absolute Time Accuracy (to UTC) & 200 ns \\
Energy Resolution @ 1 keV & 85 eV \\
Max Source Flux & 15 Crab \\
\hline
\end{tabular}
\end{center}
\end{table}

\section{Instrument Design}

\begin{figure}[h]
    \centering
    \includegraphics[width=4.5in]{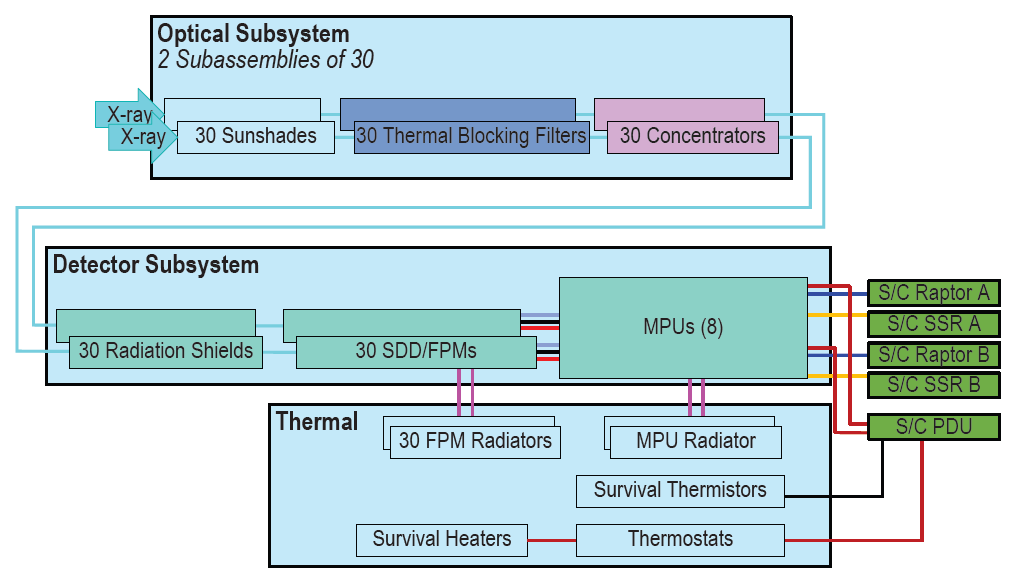}
    \caption{Functional Block Diagram of the LEMA instrument.}
    \label{fig:block}
\end{figure}

The modular design of the LEMA instrument is captured in Figure \ref{fig:block}.  Incident X-rays are collected by each XRC using just one X-ray reflection to deliver photons from the sky to the detector.  The concentrators are protected thermally by individual optical blocking filters and sunshades to avoid thermal loading. The X-rays are captured by SDDs which are surrounded by radiation shielding to protect against the space environment. Each SDD, packaged into a Focal Plane Module (FPM), shapes the charge-deposition signals from individual X-ray photons and transmits them to a Measurement and Power Unit (MPU); each MPU services up to 8 FPMs.  Both the FPMs and MPUs have individual radiators that allow for heat mitigation.  

\begin{figure}[h]
    \centering
    \includegraphics[width=\textwidth]{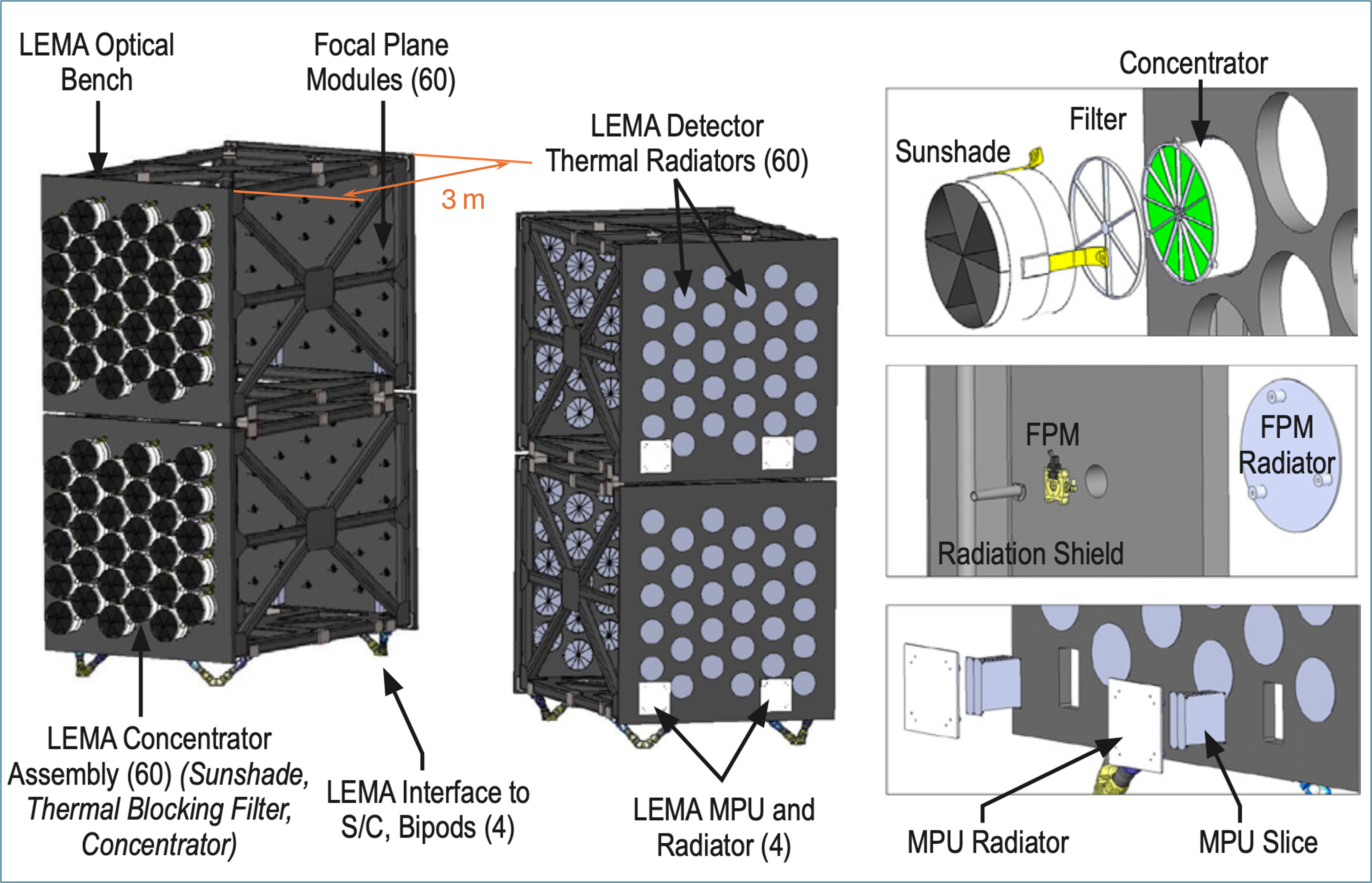}
    \caption{LEMA components and their overall layout. The instrument consists of two identical composite optical benches, \updated{with 3 m focal length,} each hosting 30 optic-detector pairs. Optical assemblies (top right) include an X-ray concentrator covered by a thermal filter and a sunshade. The detector system (lower right panels) consists of Measurement/Power Unit (MPU) ``slices,'' each of which services up to 8 focal-plane modules (FPMs, with individual radiation shielding and thermal radiators).}
    \label{fig:hardware}
\end{figure}

Figure~\ref{fig:hardware} illustrates the LEMA instrument's configuration, and the following subsections describe its individual components.

\subsection{X-Ray Concentrator (XRC) Subsystem}

The science targets for LEMA are primarily point sources, and their study does not benefit from true imaging capability. High-fidelity X-ray imaging with grazing-incidence optics requires two reflections, and thus primary and secondary mirrors. The LEMA XRCs use only a single X-ray reflection to deliver photons from the sky to the detector, with several benefits: First, the photons suffer reflection inefficiencies only once, resulting in enhanced effective area. Second, the number of optical elements required is half that of a true imager, resulting in significant mass, cost, and schedule savings. Third, integration of the instrument is much simplified, as there is no need to align primary and secondary optics.

\begin{figure}[h]
    \centering
    \includegraphics[width=\textwidth]{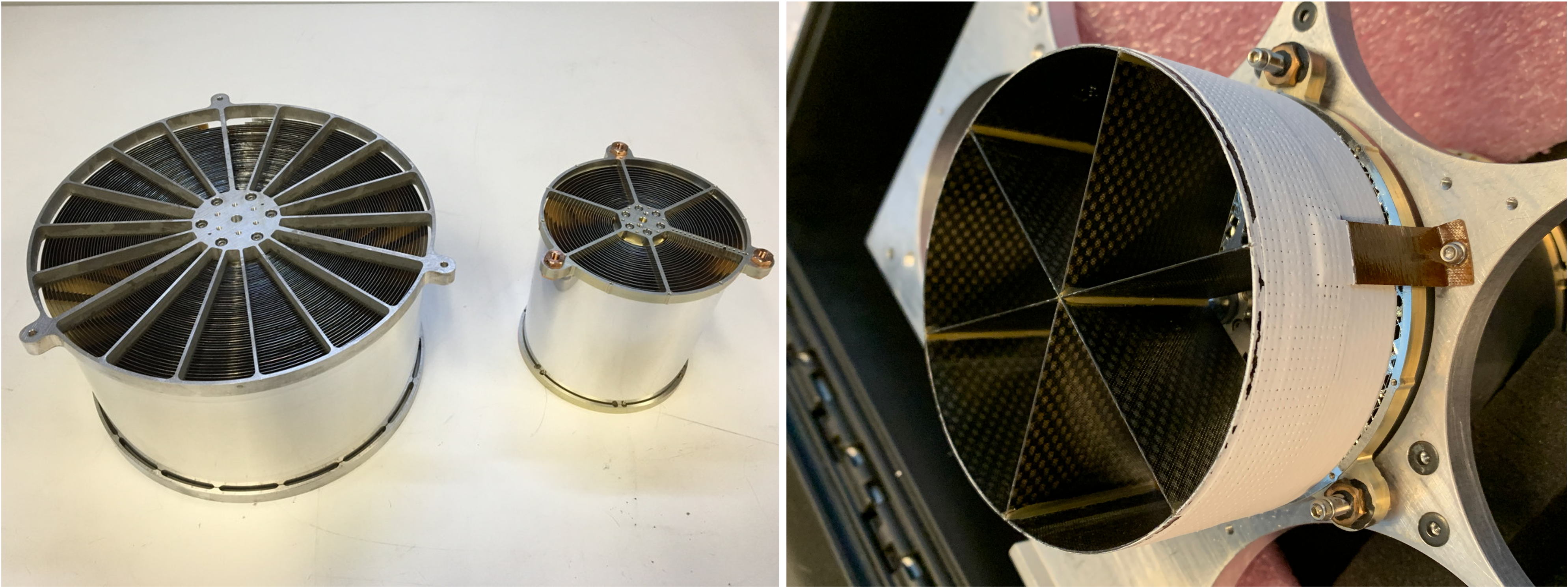}
    \caption{Left: Photo of a NICER concentrator (\updated{10.5 cm diameter}, at right) along with a scaled-up \updated{(21 cm diameter)} version, which was developed for the XACT sounding rocket program. Right: Photo of a NICER sunshade and optical blocking filter atop an X-ray concentrator mounted on a prototype optical bench (all are engineering test-unit components). }
    \label{fig:xrc}
\end{figure}

Key to meeting LEMA science objectives is collecting sufficient X-rays in the 0.2--12 keV band from targets while minimizing background. The X-ray mirror lab at GSFC has been building aluminum foil-based X-ray optics for three decades and most recently has flown 56 XRCs as part of NICER. 
\updated{The established replication process efficiently produces gold-coated foils with paraboloid figures and $\sim 5$\,\AA\ surface roughness, enabling grazing reflectivity at angles as large as 2.7$^\circ$.}
The LEMA XRCs represent a straightforward scaling of NICER's optics: increasing diameter (outermost foil) from 10.5 cm to 28 cm, increasing the number of foils within each \updated{XRC} from 24 to 107, and extending focal length from 1 m to 3 m. The performance of an intermediate configuration developed for the X-ray Advanced Concepts Testbed (XACT) sounding rocket payload---21 cm diameter, 66 foils, 2.8 m focal length (Figure \ref{fig:xrc}, left)---has been validated in X-ray beamline tests. Development of the LEMA optics thus builds on extensive experience to provide an X-ray effective area-to-mass ratio of 209 cm$^2$/kg at 1 keV, the largest yet, amounting to 66\% of the XRC's geometric area.

Each XRC front-face aperture is thermally protected by an aluminized polyimide filter that maintains dark conditions within the LEMA optical bench and minimizes optical and thermal load on the SDDs, as on NICER. The filter is mounted with the aluminized side facing outward to prevent degradation of the polyimide from atomic oxygen; it is also thermally isolated from the XRC structure and with gaps that allow venting of the interior volume. \updated{The NICER design employs a polyimide base 120 nm thick, with a 30 nm-thick Al coating, resting on a stainless-steel mesh with 93.1\% open area; a modest departure from this design, increasing thickness slightly to improve robustness against punctures and tears, may be evaluated during the STROBE-X Phase A study.} A circular sunshade, made from carbon-fiber composite material, is mounted above each XRC filter to mitigate thermal loads (Figure \ref{fig:xrc}, right). The LEMA sunshades allow the instrument to view celestial targets anywhere in the sky outside of a 45-degree exclusion zone around the Sun. The design is again direct heritage from NICER. The exterior surfaces of the sunshades are coated with AZ93 to minimize thermal energy absorption and overheating. LEMA uses a modular sunshade approach where each optic is protected by its own shade allowing a single shade to be replaced if damaged during assembly or transport. The design overlaps the spokes of the XRC inlet support structure and does not block X-ray collecting area.

\subsection{Detector System}

The STROBE-X LEMA X-ray detectors are heritage from NICER and will be procured in the same fashion: SDDs supplied by Amptek, Inc., from their standard commercial product line and integrated by MIT into a Focal Plane Module (FPM) containing a charge sensitive preamplifier board within a mechanical housing.
The SDD is contained within a hermetically sealed TO-8 package that includes a thermoelectric cooler (TEC), a thermistor, and a SAES\texttrademark{}   sintered Zirconium vacuum getter. The SDD includes an aluminized silicon nitride optical-blocking input window that shields the detector from visible light and helps to reduce thermal load on the SDD. The nominal operating environment for the detector housing within the LEMA optical bench is $\sim 20^\circ$C. Each of the 60 FPMs preamplifies the weak signals of individual X-ray photons for transmission to an MPU that can be located as far as two meters away.

Taking advantage of the pointing stability of the STROBE-X platform relative to NICER on ISS, the LEMA detector apertures are sized to provide a non-imaging (single pixel) field of view on the sky with $4'$ diameter (FWHM, compared to NICER's $6.3'$), improving sensitivity by reducing diffuse X-ray sky background as well as source confusion in crowded fields. The FPM mechanical housing derives from the NICER design, but with integral flexures to accommodate differential CTE effects at its interface with the composite LEMA optical bench. NICER utilized direct conduction to its aluminum bench for a heat path, whereas LEMA will implement a dedicated radiator at each FPM site; these localized heat paths enable the full bench to provide a stable platform from a Structural, Thermal, and Optical Performance (STOP) perspective.

Bare SDDs, like other silicon detectors, are sensitive to infrared and visible-light wavelengths. This fact is exploited during integration to precisely align each FPM to its associated XRC, a task perfected with NICER. In addition to the thermal filters covering each XRC (described above), each SDD is hermetically sealed with an aluminized \updated{(30 nm thick)} Si$_3$N$_4$ \updated{(40 nm thick)} window with 88\% open area. With these two filters in the optical path, \updated{the worst-case transmittance is $<0.05$\% at 340 nm wavelength and an order of magnitude lower at 540 nm, so that} the brightest optical counterpart LEMA can tolerate before a degradation of SDD energy resolution is V magnitude $-6.3$, orders of magnitude brighter than the visible-light emissions of any of the LEMA science targets.

\begin{figure}[h]
    \centering
    \includegraphics[width=5.5in]{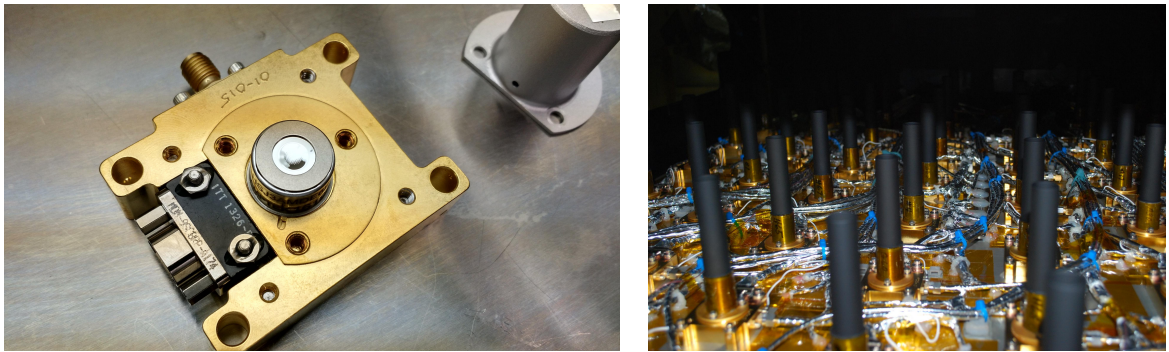}
    \caption{Left: FPM with hermetically sealed detector \updated{(central raised cylinder, 12.3 mm in diameter)} and internal TEC, with integrated electronics under the gold-colored mechanical housing. Right: NICER focal plane, \updated{where} detectors \updated{are} hidden within graded-Z cones that protect from background radiation and stray light.}
    \label{fig:fpm}
\end{figure}

\subsection{Structure}

In a departure from NICER heritage, the LEMA optical bench will be made of composite material---M55J and epoxy resin CE-3---with embedded metallic fittings, as opposed to the NICER XTI's primarily aluminum structure. This change improves on-orbit stability for LEMA by use of a low coefficient of thermal expansion (CTE) material. While X-ray missions generally have reduced alignment requirements compared to ultra-violet or visible instrumentation, STROBE-X will experience large temperature swings given its various FoV and orbital conditions, thus it is important to have the optical elements mounted to a low-CTE bench to mitigate the temperature impacts. Each half consists of two flat composite panels joined together at metallic fittings by composite tubes. 
The half is bolted and pinned together to maintain alignment throughout the mission. The design permits access to the interior of the structure for ease of detector and thermal system integration. The 
two halves are built identically to allow for cost savings during manufacturing at the supplier as well as flexibility in the assembly and integration process. 

BAE Systems, Inc., Space and Mission Systems will procure the bench and has a vast history of successful implementation for a wide variety of payload applications, including those with much more stringent stability requirements than STROBE-X. A composite bench, such as the one planned for LEMA, is individual in nature to fulfill a specific design role. As such, the common approach is to perform qualification-level testing upon completion of the bench build via thermal cycling and static load testing. This testing occurs at the supplier prior to receipt of the bench by BAE Systems, Inc., and ensures no manufacturing or latent defects will cause future potential failures.

MLI covers the external surfaces of the bench to maintain operational temperature ranges. Black Kapton covers the internal harnessing, with the remainder of the composite surfaces finished to reduce stray light impacts. These improvements are based on NICER’s operational experience and a desire to be resilient to stray light. Additionally, given the open architecture within the cavity of LEMA, internal baffling may be utilized to make LEMA even more resistant to potential optical light leaks.

\subsection{Thermal Control}

The thermal control system is a cold-biased approach using industry standards: Al radiators, MLI, and thermostats. LEMA temperature control requirements are easily met using a composite bench with isolated components compared to the more precise thermal control needed by the NICER Al bench. The wide range of solar illumination angles during STROBE-X science operations necessitate individual radiators for each detector. Sized radiator surfaces provide enough cooling for the TECs built into the detectors to keep the SDDs at $-55^\circ$C, with heaters to come on if temperatures drop too low. The thermal balance is maintained over the full range of orientations and power dissipation levels. The radiator designs accommodate CBE power $\pm$30\%\updated{, for a maximum expected value of 419 W}. The team performed detailed analyses with beta angle varying between $\pm$37.4$^\circ$, including maximum heating scenarios, multiple \updated{spacecraft} orientations, at beta 0$^\circ$ with a varying sun angle from 45$^\circ$ to 180$^\circ$. Table \ref{tab:thermal} shows the expected performance and margins against the operational temperature limits.

\begin{table}[h]
    \caption{Temperatures across all thermal cases }
    \label{tab:thermal}
    \centering
    \begin{tabular}{|l|rrr|rrr|}
    \hline
  & \multicolumn{3}{c|}{Min, $^\circ$C} & \multicolumn{3}{c|}{Max, $^\circ$C} \\
LEMA Component & Predict & FA & Margin & Predict & FA & Margin \\
\hline\hline
Detector Assembly & $-47$ & $-55$ & 8$^*$ & 5 & 15 & 5 \\
Detector Radiators & $-48$ & $-60$ & 12$^*$ & 4 & 30 & 21 \\
Concentrators & $-39$ &  $-40$ & 1$^*$ &  49 & 60 &  6 \\
MPU & $-31$ & $-35$ & 4$^*$ & 43 & 50 & 2 \\
MPU Radiator &  $-37$ & $-60$ & 23$^*$ & 15 & 60 & 40 \\
Sunshade for Concentrator & $-104$ & $-110$ & 1 & 90 & 125 & 30\\
Film for Concentrator & $-82$ & $-105$ & 18 & 72 & 240 & 163 \\
General Structure & $-52$ & $-60$ & 3 & 16 & 60 & 39 \\
\hline
    \end{tabular}

\small Flight allowable (FA) holds $\pm 5^{\circ}$C uncertainty \\
\small Margin is reported as $^\circ$C beyond uncertainty \\     
\small $^*$except cold side heater-controlled items where 30\% heater margin overrides 
\end{table}

\subsection{MPU Electronics and Software}

 Groupings of 7 or 8 FPMs are connected to a single MPU. Each MPU controls the FPM TECs, provides a bias voltage for the FPM detectors, and receives the amplified signals from the detectors. The MPU has a fast channel in parallel with a slow channel for each FPM detector. Both of the LEMA optical benches host 4 MPU slices, each with 8 available channels to accommodate all 30 of the detectors on each bench, with two spare channels.

Each MPU slice has an RS422 interface to receive commands and send digital data back to the spacecraft, as well as a Pulse Per Second (PPS) time-tick signal input; this reference is used to convert the internal MPU processor cycle count, with which photons are initially time-stamped, to a time value based on the spacecraft's GPS receiver, with very low absolute uncertainty, demonstrated to be better than 100 ns (1 $\sigma$) with NICER. A redundant LVDS High Speed Serial (HSS) interface to the \updated{spacecraft} transmits the science data\updated{, with an average rate less than 3 Mbit/s,} directly to a \updated{1 Tbit-capacity} solid-state recorder (SSR) \updated{that serves all three STROBE-X instruments}.

Each MPU slice monitors the signal line from up to 8 FPMs, triggering on any  change that exceeds a commandable threshold. Once an FPM channel triggers, the event is time-tagged and the analog level is held in a queue to await further processing while the channel is disabled from additional triggering. 
\updated{The time interval during which event processing disables a channel is recorded and packaged with event telemetry, providing precise measurement of deadtime for each detector.}
Both slow and fast shaping circuits filter the event for high frequency noise. The slower channel, with a much longer shaping time ($\sim 300$ ns), provides better energy resolution of the detected X-rays, which enables the LEMA spectroscopic science objectives and a cleaner selection against background events. The fast channel includes higher frequency components that produce better time-tag precision ($\sim 30$ ns). For events that trigger both the slow and fast chains, the fast channel provides the event time. 

LEMA implements two low-risk modifications to the MPU: an upgrade from an obsolete plastic microcontroller to a rad-tolerant FPGA (RTG4 reprogrammable), and enhanced SDD reset control via charge monitoring in the MPU to control FPM resets so they do not trigger as “undershoot” events on the FPM signal line. FPM resets occur because of the need to flush charge from each FPM's accumulation capacitor when it reaches its capacity. GSFC has funded and largely completed the FPGA and reset algorithm updates for other instrumentation projects, so they will be fully validated and available before STROBE-X Phase B begins.

\section{Instrument Performance and Calibration}

Based on physical models, simulations (e.g., ray tracing), and laboratory measurements---all validated by on-orbit calibration of the NICER XTI---the LEMA instrument's expected performance is well understood. 

Following fabrication of $\sim 70$ XRCs (exceeding the 60 units needed for flight), testing at GSFC's 600-meter X-ray beamline\cite{2013SPIE.8861E..1MB} establishes their performance and ranks the XRCs according to on-axis effective area and sharpness of concentrated flux (e.g., percentage captured within the LEMA detector aperture\updated{, typically 88\%--94\%}). The best XRCs\updated{, ranked by the product of peak effective area and percentage of flux within the aperture,} are selected for flight. Additional ground measurements include X-ray reflectivity for a sample of XRC foils, energy-dependent X-ray absorption of the thermal filters, and XRC throughput in visible light both upon fabrication and for alignment with FPM detectors during integration within the optical bench. \updated{(Within the observatory-level pointing and alignment budget, the LEMA optic-detector boresights are allocated maximum misalignments of 25 arcseconds in  tip-tilt excursions.)} The results of these measurements inform on-orbit calibration using celestial X-ray sources. Data from scans of the LEMA pointing axis across the Crab Nebula and other well-characterized targets\cite{2017AJ....153....2M} feed into a comprehensive model that simultaneously determines the relative co-alignments of the 60 XRCs and their individual energy-dependent effective areas as a function of off-axis angle (i.e., vignetting \updated{due to the combination of single-reflection coma and the detector aperture}), down to the contributions of individual foils. 

\begin{figure}[h]
    \centering
    \includegraphics[width=4.25in]{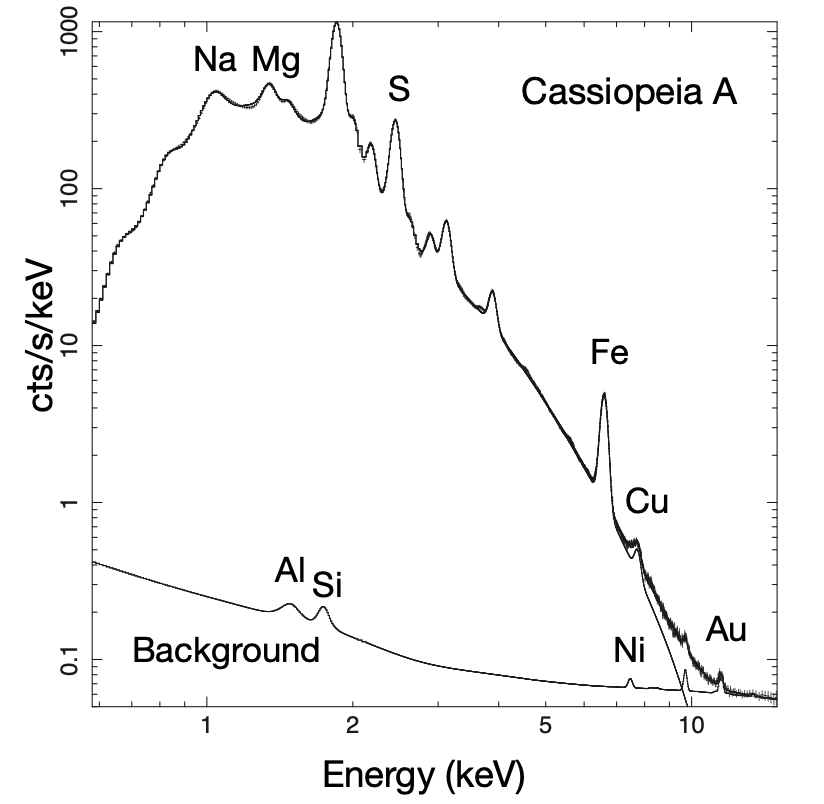}
    \caption{Spectrum of the supernova remnant Cas A, a widely used calibration standard, acquired with NICER. Characteristic emission lines from many elements are evident in the total spectrum (source + background; stepped curve), with fluorescence features from XTI hardware contributing to the background. The smooth curves, for background and the Cas A source spectrum (overplotted on the data) separately, are models derived with SCORPEON. The STROBE-X LEMA will provide the same source spectrum, with less background, in less than 1/8th the total NICER exposure.}
    \label{fig:spec}
\end{figure}

FPMs and MPUs are initially characterized in a laboratory setting\cite{2016SPIE.9905E..1IP}, to assess energy resolution, detection threshold efficiencies at low energies for both the slow and fast signal chains, and SDD reset rates. To the extent possible, SDDs are procured so that the active silicon for the entire lot originates from a single wafer, with identical surface passivation and other properties. The best-performing SDDs are selected for flight. A variety of measurements---such as X-ray absorption curves for the detector windows, gain functions to map charge-sensitive preamplifier voltages to photon energies, and ``ballistic deficit'' delays between the slow and fast chains that enable \updated{discrimination} of high-energy radiation background---may be made using synchrotron beamline facilities. Finally, any temperature dependence of FPM-MPU performance is assessed during thermal-vac testing of the integrated instrument, and overall time delays between photon detection and GPS time stamps are measured using a pulsed X-ray source and GPS radio-frequency simulators coupled to the STROBE-X GPS receiver. On-orbit calibration of the detector system proceeds through observations of calibration standards,\cite{2017AJ....153....2M} including the Cas A supernova remnant (Fig.~\ref{fig:spec}); radiation-zone passages also produce fluorescence lines of aluminum, gold, and nickel that can be used to track any gain changes over time. 

Instrument response files for use in data analysis draw from the ground and on-orbit measurements described above, together with physical models such as that of Scholze \& Procop\cite{2009XRS....38..312S} tailored for the LEMA SDD configuration. A sophisticated model of instrument and cosmic background components, with heritage from NICER's SCORPEON\cite{url:scorpeon}
but substantially simplified thanks to the low-inclination orbit of STROBE-X, provides the capability to both predict and account for background terms during data analysis. \updated{Specifically, differences in the relevance of various background contributors for NICER and LEMA scale according to 1) the instrument collecting area, for the diffuse cosmic X-ray background, and 2) the number of detectors and the operational particle and gamma-ray environment, for the radiation component. The astrophysical X-ray background dominates at energies below 2 keV; LEMA's much higher collecting area is partially offset by its smaller aperture to the sky, resulting in a relative increase in raw background rate by a factor $\sim 3$. The ambient radiation background rate, spectrally flat across the LEMA band, is comparable to the best conditions seen by NICER at low geomagnetic latitudes, largely immune to the ``trapped electron'' and ``precipitating electron'' components of the SCORPEON model near the Earth's auroral zones.}

\begin{figure}[h]
    \centering
    \includegraphics[width=4.25in]{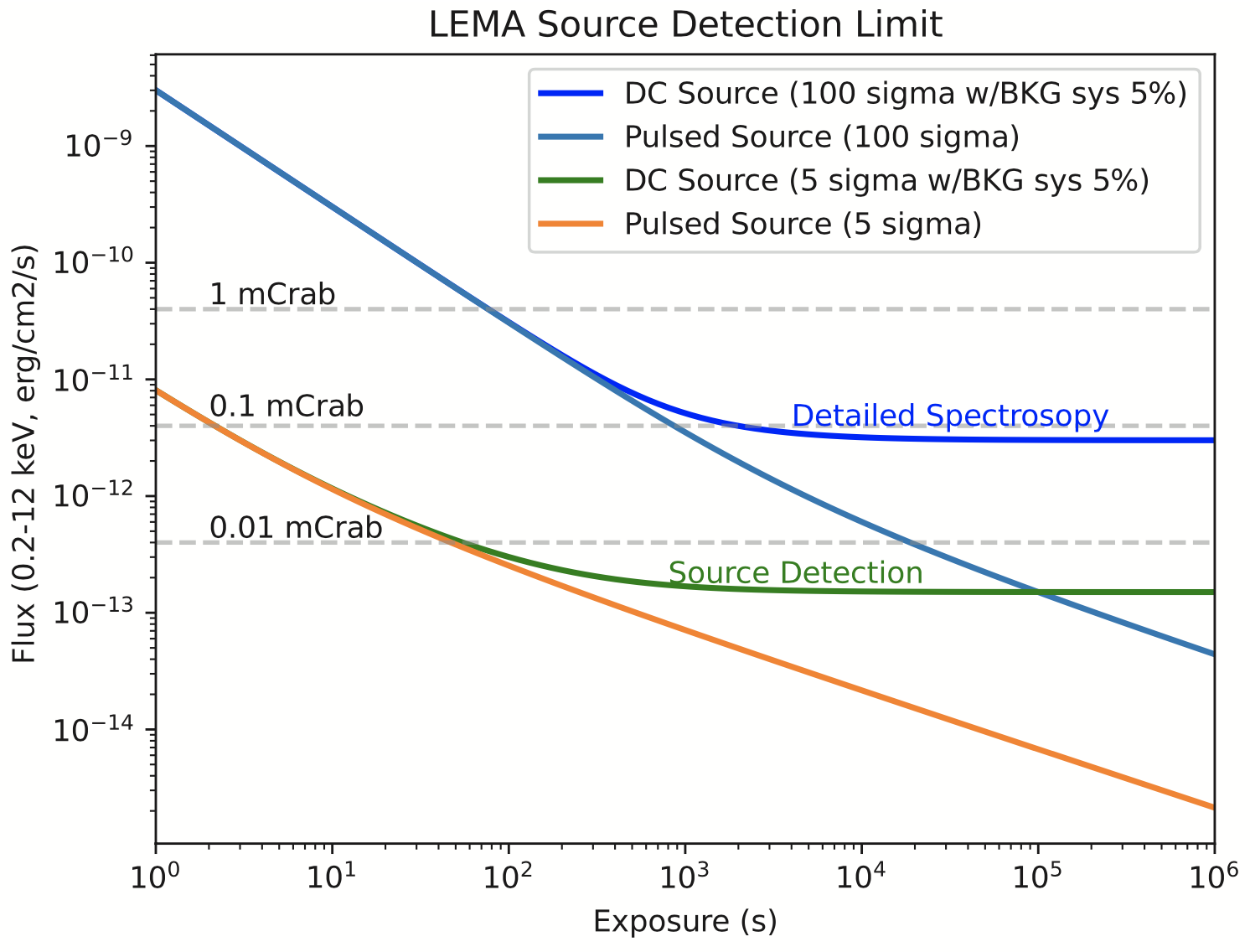}
    \caption{Expected sensitivity of LEMA to steady (DC) and pulsed sources of broadband X-rays, as a function of exposure time and at two levels of statistical significance: 100$\sigma$ and 5$\sigma$, including systematic uncertainty in knowledge of the background at the 5\% level for unpulsed sources.}
    \label{fig:sens}
\end{figure}

Taken together, the expected performance of the instrument's components yields reliable predictions for LEMA's sensitivity to X-ray sources emitting across the 0.2--12 keV band. As shown in Fig.~\ref{fig:sens}, for example, LEMA will be capable of detecting steady sources as faint as 5~$\mu$Crab in flux in exposures of just 1 ks.

\section{Technology Readiness Level (TRL) and Development Plan}

Much of the technology and design of the LEMA instrument is comfortably at TRL 6 or higher (functional tests in a space-like environment, with TRL 9 representing the highest level of maturity), benefiting from direct heritage from the NICER instrument---the concentrators, detector assemblies, and electronics all stem from NICER heritage. The concentrators are within the manufacturing parameters of past hardware built by Goddard’s X-ray Mirror Lab. The detector assemblies have been proven and flown on NICER; slight modifications will take advantage of their use on a free-flyer mission. The electronics gain an upgrade, from an obsolete plastic microcontroller to a rad-tolerant FPGA, yet maintain the majority of their architecture. The optical bench, while new to LEMA, is built on decades of engineering experience and can be implemented with low risk. It is noted that the LEMA Lead is currently funded by planetary science programs to build and qualify FPGA-based MPUs and firmware in the coming year, gaining further experience with the LEMA modifications available before STROBE-X Phase B.

\section{Instrument Operations}


With the narrowest FOV among the three instruments onboard STROBE-X, LEMA drives pointing requirements. For this reason, the STROBE-X star trackers, which provide feedback for accurate pointing, are mounted on the LEMA optical bench. Boresight offsets between the XRCs and star trackers are assessed during integration, and fully calibrated (at the arcsecond level) during on-orbit checkout. For science operations, routine commanding at the spacecraft level reorients STROBE-X for observations---with both LEMA and HEMA---of each successive science target. The LEMA detectors continue to operate during slews in orbit-night, but collection of FPM event data is paused while the spacecraft slews in orbit-day, to reduce telemetry loading due to high detector reset rates if the slew path passes near the Sun. FPMs and MPUs remain enabled during passages through the South Atlantic Anomaly (SAA)---the small size of the SDDs and their radiation shielding result in relatively low event rates that are easily accommodated by the detector readout system.    

The LEMA detector electronics offer a single, all-encompassing data acquisition mode: all events (X-rays as well as any energy deposition in the SDDs due to particle or photon background radiation) are registered and packaged into telemetry. This approach, taking advantage of ample available telemetry bandwidth, simplifies both operations (no time-consuming instrument configuration set-up is needed prior to every observation) and data pipeline-processing and analysis. The only configuration choice that may be desirable during routine science operations is whether to disable a certain number of FPMs to accommodate extremely bright targets ($> 15$ Crab in flux across the LEMA band), those that could overwhelm the processing of events within the MPU slices or saturate the internal telemetry connections between the MPUs and the main electronics box. These instances are expected to be rare (but are scientifically valuable). \updated{Even at rates as high as 15 Crab, testing of NICER SDDs at the BESSY synchrotron facility\cite{2016SPIE.9905E..1IP} suggests that pile-up---the incidence of more than one photon within one pulse shaping time in the detector readout electronics---will not exceed the few-percent level for the slow channel and $<1$\% for the fast channel.}

On long timescales---every few years---LEMA operations accommodate several days of room-temperature annealing of SDDs, to reduce noise from increased dark current that may result from radiation damage; this is accomplished simply by turning off the TEC cooler within the SDD, allowing it to warm up to the radiator temperature. Annealing is performed for a few detectors at a time while the rest of the instrument continues to acquire science data, a benefit of the modularity and redundancy of the LEMA design. The lower orbital inclination of STROBE-X compared to NICER on ISS suggests that the incidence and rate of radiation damage that causes growth in dark current will be substantially less than the already low levels seen in NICER.

\section{Conclusion}

The Low Energy Modular Array provides an order of magnitude increase in sensitivity over previous instruments, with modular design, low background, and huge dynamic range in target intensity. The hardware is an evolution of the NICER instrument currently performing science observations onboard the International Space Station, providing a massive increase in sensitivity, reduced background, and unobstructed sky visibility available on a free-flying mission. With LEMA working in concert with the HEMA and WFM instruments, STROBE-X serves as a game-changing observatory for time-domain and multi-messenger (TDAMM) astrophysics and high-throughput time-dependent spectroscopy.

\appendix    

\subsection*{Disclosures}
The authors have no potential conflicts of interest to disclose.

\subsection* {Code and Data Availability} 
Data sharing is not applicable to this article, as no new data were created or analyzed.

\subsection*{Acknowledgments}

Portions of this work performed at NRL were supported by ONR 6.1 basic research funding.


\bibliography{report}   
\bibliographystyle{spiejour}   


\vspace{2ex}\noindent\textbf{Keith Gendreau} is an astrophysicist at NASA's Goddard Space Flight Center (GSFC).
\vspace{2ex}

\noindent\textbf{Dominic Maes} is a structural engineer at BAE Systems, Inc. Space and Missions Systems with 11 years of experience in the development of space flight instruments, including the \textit{Roman} Space Telescope and TEMPO.
\vspace{2ex}

\noindent\textbf{Ronald A Remillard} (PhD., 1985, MIT) is an astrophysicist at the MIT-Kavli Institute for Astrophysics and Space Research. His primary research interest is the optical and X-ray emission from black hole binary systems. He has been the science lead for instruments such as the All-Sky Monitor on the \textit{Rossi} X-ray Timing Explorer and the detector team at MIT for the NICER mission on the International Space Station.
\vspace{2ex}

\noindent\textbf{Paul Ray} is an astrophysicist at the U.S. Naval Research Laboratory. He received his A.B. in physics from the University of California, Berkeley in 1989 and his Ph.D. in physics from Caltech in 1995. His current research interests include multiwavelength studies of pulsars and other neutron star systems and X-ray instrument and mission development. He is a member of AAS, IAU, and Sigma Xi.
\vspace{2ex}

\noindent\textbf{Zaven Arzoumanian} is an astrophysicist at NASA's GSFC.
\vspace{2ex}

\noindent Biographies and photographs of the other authors are not available.


\end{document}